# Examining UK drill music through sentiment trajectory analysis

Bennett Kleinberg[1,2] and Paul McFarlane[1,3]

[1] Department of Security and Crime Science, University College London
[2] Dawes Centre for Future Crime, University College London
[3] Centre for Global City Policing, University College London

**Abstract** This paper presents how techniques from natural language processing can be used to examine the sentiment trajectories of gang-related drill music in the United Kingdom (UK). This work is important because key public figures are loosely making controversial linkages between drill music and recent escalations in youth violence in London. Thus, this paper examines the dynamic use of sentiment in gang-related drill music lyrics. The findings suggest two distinct sentiment use patterns and statistical analyses revealed that lyrics with a markedly positive tone attract more views and engagement on YouTube than negative ones. Our work provides the first empirical insights into the language use of London drill music, and it can, therefore, be used in future studies and by policymakers to help understand the alleged drill-gang nexus.

**Keywords:** social media, sentiment analysis, drill music, gang crime, lyrics

## 1 Introduction

Gang-related violence is increasing in the UK. While there have been noteworthy reductions in many crime types, since 2017, there has been an alarming rise in serious violence, particularly gun and knife crimes involving young people (Office for National Statistics, 2018). Tackling this increase in serious violence is now more complicated because current research contends that street gang-rivalries, which often result in violence, have moved to online social media platforms. In these cases, social media is used to communicate hostile intentions and deliberately antagonise opposing street gangs (Décary-Hétu & Morselli, 2011; Pinkney & Robinson-Edwards, 2018; Pyrooz, Decker, & Jr, 2015).

At the same time, as the increase in violence in the UK, social media platforms such as YouTube and Instagram have witnessed significant growth in drill music videos uploaded by London street gangs (Dearden, 2018). Drill music, a subgenre of Chicago hip-hop, has spread to major cities in the UK and the videos uploaded to social media platforms receive considerable attention in the form of views and comments. In the UK, key public figures are concerned and hint at a potential relationship between the two. They have sought to develop a narrative linking social media, particularly, platforms that host drill music videos, to increases in gang violence involving young people. For instance, the most senior police officer in the UK, Commissioner

Cressida Dick, has publicly criticised how drill music is used to incite inter gang-violence. She claims drill music lyrics are:

> " […] about glamorising serious violence: murder, stabbings […] they describe the stabbings in great detail, joy and excitement […] in London, we have gangs who make drill music, and, in those videos, they taunt each other. They say what they are going to do to each other and specifically what they are going to do to who" (Waterson, 2018).

What we do know, unique to urban street gangs, is that drill music lyrics are also used to express narratives and sentiment about perceived societal grievances (Pinkney & Robinson-Edwards, 2018). For instance, the song *Tripidy Trap* by *Rushy* is just one illustration of how drill lyrics are used to describe the harsh realities of living in poverty-stricken inner cities in the UK. In this song, *Rushy* raps about not liking life and the need to be involved in drug dealing to earn a living (*Rushy—Trippidy Trap*, 2019). Moreover, while supporters disagree with the police blaming drill music, there has been an increase in young people involved in creating and producing drill music, being killed in the UK (Cobain, 2018). For example, shortly before his alleged gang-related death, the drill music rapper, *Incognito*, from the London Street gang *Moscow 17* gave an interview (see Mardean, 2018). He stated in the interview that "you've got to put your hands up and say, drill music does influence it" when talking about the nexus between drill music and violence.

Nevertheless, research on the drill music-gang violence nexus is scarce. Although some research does suggest that online social media activity and gang-related violence are correlated (Johnson & Schell-Busey, 2016), there is currently no direct evidence indicating a relationship between drill music lyrics and gang violence. One particular challenge is the domain-specific language that characterises drill music lyrics. This paper provides the first glimpse at the drill music phenomenon by examining the lyrics through an intra-textual sentiment approach using techniques from natural language processing. The approach here employs an automated analytical workflow potentially suitable with understanding the problem at scale.

## 1.1 Sentiment trajectories

Sentiment analysis is concerned with measuring and modelling the valence of a text and has been proven useful in identifying potentially radical authors in online forums (Scrivens, Davies, Frank, & Mei, 2015), measuring attitudes towards different entities in student evaluations (Welch & Mihalcea, 2016), and many more. Recently, a shift has been proposed away from static approaches (i.e. one sentiment score per text) to those that use smaller units of analysis (e.g., sentences, Rinker, 2018). The underlying idea is that sentiment is a linguistic construct that is used dynamically within a text – a characteristic that is obscured by static approaches (Gao, Jockers, Laudun, & Tangherlini, 2016; Kleinberg, Mozes, & Van der Vegt, 2018; Reagan, Mitchell, Kiley, Danforth, & Dodds, 2016). The notion of automatic, dynamic sentiment extraction has been pioneered in work on novel arcs (Jockers, 2015b; Reagan et al., 2016) and was recently adopted for the study of public speeches (Tanveer, Samrose, Baten, & Hoque, 2018), YouTube vlogs (Kleinberg et al., 2018), and left- and right-wing news channels (Soldner et al., 2019).

A fundamental limitation of noisy text (e.g., Tweets, vlogs, lyrics) is that they do not adhere to common English sentence structures and punctuation and are, therefore, not suitable for sentenced-based dynamic approaches. An alternative method has been proposed that uses context windows around sentiment terms so that if functions agnostic to sentence structure (Kleinberg et al., 2018; Soldner et al., 2019). The current work adopts such an approach because drill music lyrics are closer to noisy user-generated text than to standard written English.

### 1.2 Domain-specific language challenge

Aside from the noisiness of drill music lyrics, the idiomatic language poses an additional challenge. Previous work on a social-media escalation of Chicago-based gang members, for example, acknowledged that problem and built a unique Chicago gang language phrasebook (Blevins et al., 2016). Others used word embeddings to circumvent the problem of highly context-dependent language used by gang members of Twitter, and hence often unknown terms for automated analysis (Wijeratne, Balasuriya, Doran, & Sheth, 2016), while yet other work used slang translation approaches to normalise idiomatic language into standard English (Han & Baldwin, 2011). The current work employs two types of sentiment measurements: a standard, common English sentiment lexicon, and a slang-specific one (SlangSD; Wu, Morstatter, & Liu, 2018).

### 1.3 Aims

We present a sentiment trajectory analysis of the London drill music scene using both a standard and a slang-specific sentiment approach and seek to shed light on how drill music employs sentiment. By doing so, we offer the first empirical investigation into the language use of a highly-debated, current social media phenomenon.

## 2 Method

### 2.1 Data

We aimed to create a comprehensive corpus of the London drill music scene. First, we identified drill music artists using two open-source lists[1] and selected those that are reportedly associated with London street gangs (*n* = 127). Second, we extracted the URLs of the lyrics of all artists (*n* = 744 songs) from genius.com and scraped the lyrics through parsehub.com. Third, we cross-referenced the songs with videos posted on YouTube and retrieved publicly available metadata (i.e. view count, (dis)like count, comment count) for each video through YouTube's API. After merging the data, pre-processing and duplicate removal, the final corpus consisted of 550 songs with lyrics and YouTube metadata from 105 different artists (Table 1).

Table 1. Corpus descriptive statistics.

| Variable | Mean (SD) [99% CI of mean] |
|---|---|
| Tokens | 644.06 (218.51) [620.02; 668.10] |
| Type-token ratio | 0.43 (0.09) [0.42; 0.44] |
| Comments | 418.64 (734.16) [337.87; 499.40] |
| Likes | 7632.07 (13837.58) [6109.77; 9154.36] |
| Dislikes | 197.95 (425.68) [151.12; 244.78] |
| Days active | 572.27 (387.95) [529.59; 614.94] |

---

[1] https://www.last.fm/tag/uk+drill/artists; https://www.dummymag.com/10-best/10-of-the-best-uk-drill-tracks-according-to-67/

## 2.2 Sentiment trajectories

This study's primary aim was to model the sentiment using an approach that accounts for dynamics within each lyric. We used two sentiment lexicons: the *Jockers & Rinker Polarity Lookup Table* (Rinker, 2018a) and the *SlangSD* lexicon (Wu et al., 2018), both of which are contained in the lexicon *R* package (Rinker, 2018a). While the former offers a standard sentiment lexicon (11,710 terms), the *SlangSD* lexicon was explicitly designed to capture the sentiment of slang or highly domain-specific language (48,277 terms). Hence, the slang sentiment lexicon measures terms that the standard sentiment lexicon does not contain (e.g., "bloods") and measures overlapping terms differently (e.g., "sex", measured as 0.10 in the standard and as -0.50 in the slang sentiment).

For each lyric, we extracted the raw sentiment trajectory accounting for valence-shifters (i.e. negators, amplifiers, de-amplifiers, adversative conjunctions) using a word window around each sentiment term (see Kleinberg et al., 2018; Soldner et al., 2019)[2]. Specifically, for each sentiment match, a sentiment context window of +/- 3 words was constructed in which any valence shifters were identified and the sentiment corrected accordingly (e.g., if a negation preceded a sentiment in the context window, the valence was shifted to the opposite). This procedure produces a sparse vector of the length of the lyrics with zeros for non-sentiment matches and corrected sentiment values. To obtain comparable representations of the sentiments, we forced each lyric to a standardized narrative time ranging from 1 to 100. The standardization was achieved by applying a discrete cosine transformation to the sparse vector and transform it to a new vector of length 100 for each lyric (Jockers, 2015a; Kleinberg et al., 2018; Soldner et al., 2019). The length of the transformed vector was chosen as 100 so that it is interpretable as percentages of lyrics progression (i.e. each of the 100 bins corresponds to 1% of the length of the lyrics).

Figure 1 shows two example trajectories from our drill music corpus. The first one (*Moscow 17 – Blessed*) shows a sentiment development that starts negative and becomes positive with a peak between 60-80% of the narrative progression. The lyrics here are 474 words long, so 60-80% of the progression corresponds to the lyrics between words 284 and 379. The text ends slightly negatively below zero.

For the second example (*Headie One – Different Sorts*), the lyrics remain in the negative sentiment area almost throughout the entire song. In the beginning, the negative is only slightly below zero and becomes marginally positive at a standardized progression of ca. 40%, before it develops to a stronger negative sentiment between 60-80% of the lyrics' progression. Note that the length of the lyrics is different in this case: the song *Different Sorts* has 962 words so that the standardised time of 60-80% corresponds to the lyrics between 577 and 770 words.

## 2.3 Analysis plan

We were interested in assessing whether there are distinct trajectory shapes in the standard and slang sentiment use. We applied unsupervised learning (*k*-means clustering) on the two sentiment vectors (standard sentiment and slang sentiment) and used the elbow method and Silhouette method to decide for a number of clusters *k*. Further, we examined the associations between the resulting clusters to understand interrelationships between the two sentiment types. Finally, we looked at the relationship between the popularity of drill music videos on YouTube and the sentiment clusters using negative binomial regression.

---

[2] Note that we do not force the sentiment to a predefined scale, contrary to other work (Gao, Jockers, Laudun, & Tangherlini, 2016; Kleinberg, Mozes, & Van der Vegt, 2018).

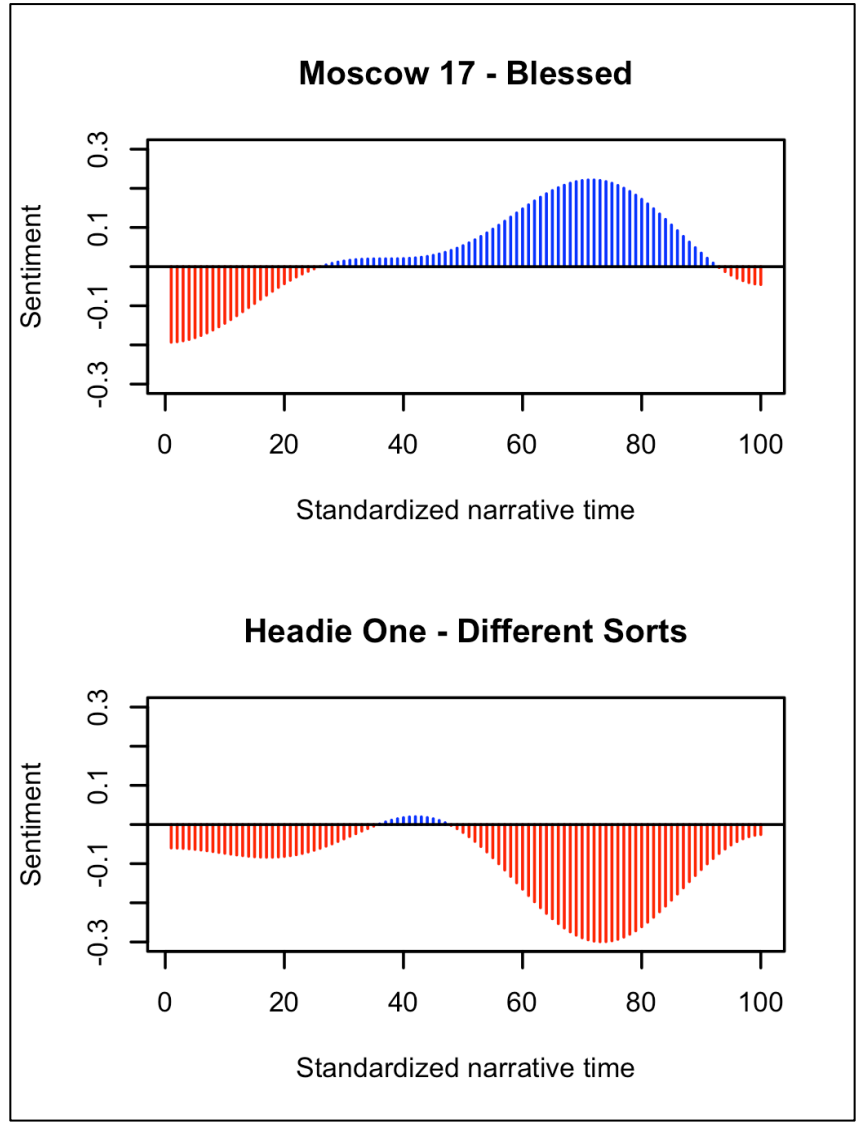

Figure 1: Two example sentiment trajectories.

## 3 Results

### 3.1 Standard sentiment trajectories

Preliminary analyses suggested that the ideal number of clusters $k$ was $k=2$. Figure 2 shows the average standard sentiment shape for each of the two clusters. The averaging procedure took each of the 100 bins on the narrative time axis and calculated the mean value of all lyrics in that cluster, resulting in one average value per bin. The resulting average of all 100 bins represents the average shape per cluster (for median aggregation plots, see Appendix A and B). The trajectory shapes suggest that one cluster captured drill lyrics that were positive in sentiment throughout with a pronounced positive start and end. We call this cluster the "positive cluster", and it captured 53% of all drill lyrics ($n = 289$). The 99% confidence interval (CI) of the mean sentiment value per narrative time bin, as well as the population standard deviations, indicate that the sentiment is not conclusively positive. The 99% CI shows that the sentiment moves below the zero line for a narrative time between 20-80%.

Conversely, the second cluster, called the "negative cluster" (47% of all lyrics, $n = 261$), captured drill lyrics that remained negative throughout although the standard deviations show that there is some fluctuation that goes into the positive sentiment area. The shapes overall indicate a stronger negative sentiment at the start and end. These two clusters (positive and negative) are like mirror images and seem to represent two classes of drill lyrics.

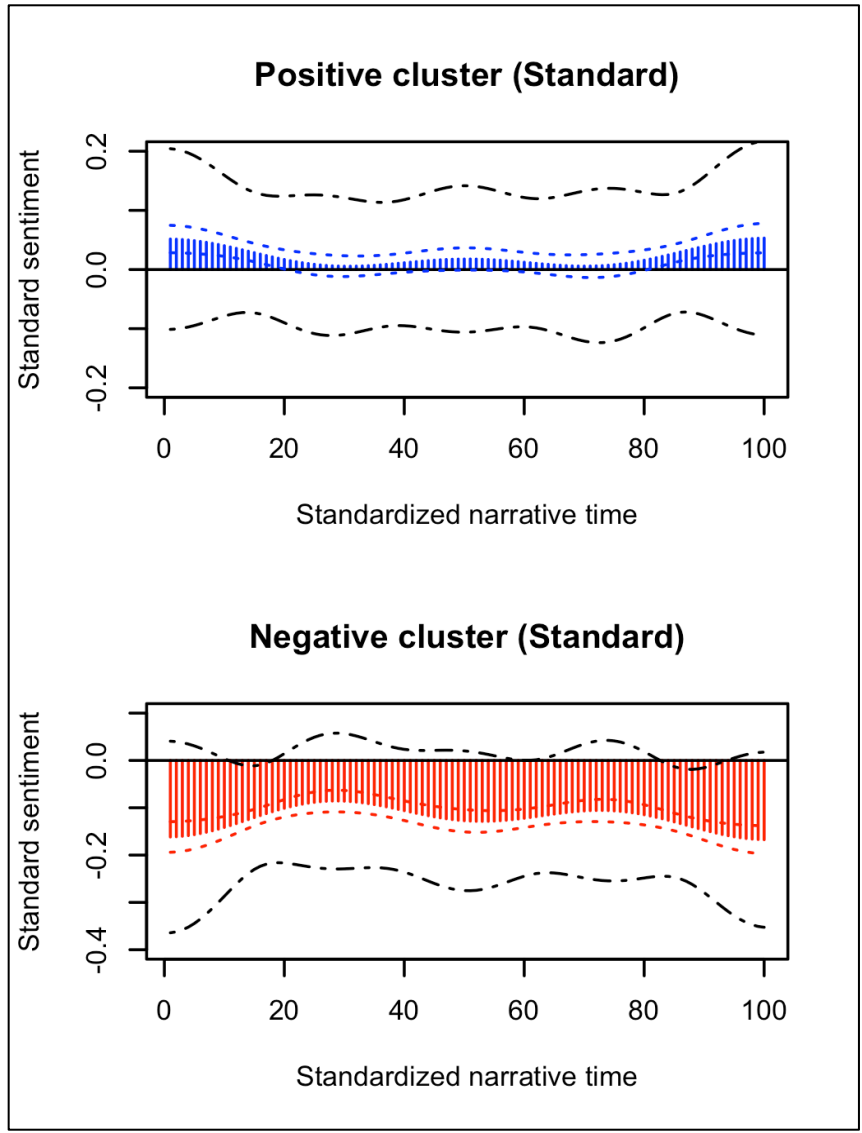

Figure 2: Mean sentiment clusters using the standard sentiment lexicon
(red/blue dashed lines = 99% conf. interval, black dashed lines = +/- 1 SD)

### 3.2  Slang sentiment trajectories

For slang sentiment trajectories too, preliminary analyses suggested $k=2$ for the $k$-means model. We followed the same procedure as for the standard sentiment vectors: the slang sentiment trajectories were averaged by cluster membership, resulting in two aggregated shapes (Fig. 3).

Both clusters represent lyrics with a consistently negative sentiment; one with a sentiment ranging between -0.05 and -0.10 ("negative cluster 1", 70% of all lyrics, *n* = 386), the other with a stronger negative sentiment between -0.20 and -0.30 ("negative cluster 2", 30% of all lyrics, *n* = 164). The standard deviations suggest some fluctuation above the zero line for the first negative sentiment cluster while the sentiment remains in the negative area entirely for the second cluster.

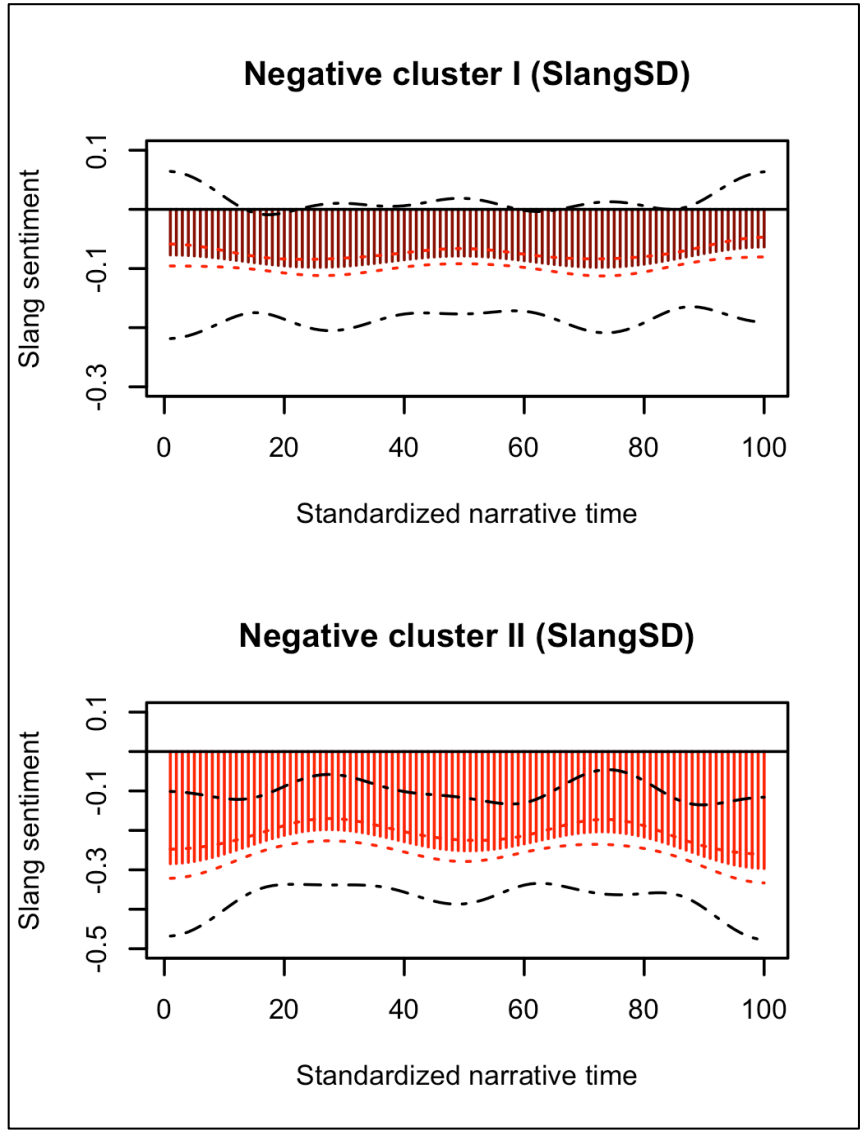

Figure 3: Mean sentiment clusters using the SlangSD sentiment lexicon

(red dashed lines = 99% conf. interval, black dashed lines = +/- 1 SD)

### 3.3 Cross-cluster analysis

The standard sentiment and the slang sentiment method may capture the same linguistic construct, with the latter merely providing a more accurate measurement. If that were the case, we would expect a significant association between cluster membership of the standard and slang sentiment. A Chi-square test on cross-cluster membership (Table 2) did not reveal a significant association, $X^2(1) = 2.05$, $p = .152$. These results suggest that cluster membership of one sentiment type is independent of membership of the other.

Table 2: Cross-cluster distribution

|  | **Negative slang I** | **Negative slang II** | **TOTAL** |
|---|---|---|---|
| **Positive standard** | 211 | 78 | 289 |
| **Negative standard** | 175 | 86 | 261 |
| **TOTAL** | 386 | 164 | 550 |

### 3.4 Video popularity and engagement

Lastly, we examined whether the clusters of either sentiment type were related to the popularity of and engagement with the drill music video on YouTube. Video popularity was operationalised with two indices: (i) the number of views per 100 days and (ii) the overall community engagement (= the sum of likes/100 days, dislikes/100 days, comments/100 days). For each video popularity index, we ran a 2 (standard sentiment: positive vs negative) by 2 (slang sentiment: negative I vs negative II) negative binomial regression model (since the counts of views/100days and community engagement were over-dispersed).

The regression analyses (Table 3) revealed that for the view count model, only the cluster membership of the standard sentiment was a significant predictor. The views per 100 days were, on average, almost twice as high ($\frac{1}{e^{-0.64}} = 1.89$) for lyrics in the positive cluster ($M$ = 236,706.90, $SD$ = 592,135.80) compared to the negative cluster ($M$ = 140,090.00, $SD$ = 315,327.80). Similarly, the video engagement counts of drill lyrics in the positive sentiment cluster ($M$ = 2,745.35, $SD$ = 7,181.17) were more than twice as high ($\frac{1}{e^{-0.70}} = 2.01$) as in the negative cluster ($M$ = 1,644.16, $SD$ = 3,139.25).

Taken together, these findings suggest that drill music tracks with lyrics in the positive standard sentiment cluster attracted considerably more views and elicited more community engagement than lyrics in the markedly negative cluster. There were no effects for the slang sentiment clusters nor interaction effects between the two.

Table 3. Coefficients of negative binomial models.

|  | **Standard sentiment** | **Slang sentiment** | **Interaction** |
|---|---|---|---|
| View count model | -0.64 (0.17)* | 0.15 (0.22) | 0.28 (0.31) |
| Engagement model | -0.70 (0.16)* | 0.05 (0.21) | 0.46 (0.29) |

Note. The reference group was "negative" for standard and "very negative" for slang sentiment. * = sign. at $p$ < .001. Coefficients are not anti-logged.

### 4 Discussion

This paper examined how sentiment is used dynamically in drill music associated with the London gang scene. Since the vocabulary of drill music differs from standard English (here: on average 18.64% of words were not among the 10k most frequently used English words), we measured the sentiment of lyrics in two ways. For both a standard and a slang-specific sentiment measure, there was evidence of clustering in two distinct sentiment trajectories each. While there was a positive and a mirrored, negative sentiment shape cluster for the standard sentiment, the slang-specific sentiment resulted in two negative clusters with different intensities of negativity. We found no evidence that cluster membership for both sentiment types is associated with one another, suggesting that they do indeed measure different kinds of sentiment.

An important channel of communication for street gangs is social media, so we examined the popularity of drill music videos on YouTube. Our findings indicated that the positive sentiment trajectory attracted almost twice as many views and resulted in twice as much engagement with the content as drill videos that used a negative sentiment style. These findings need further corroboration but might in the future help practitioners (e.g., the police) to better grasp the relationship between potentially violent messages and drill music popularity (for a related comments, see McFarlane, 2019).

The nexus between gang violence and drill music is yet to be corroborated by rigorous research. However, an approach like the one presented here could be a starting point in studying the content released by street gangs on YouTube. An avenue for future work could lie in examining not only the actual content but also the engagement with the content. Possibly, the tension between street gangs could be measured through comments to music videos, which in turn could be explored as a means to forecast real-world gang rivalry escalations. Suchlike studies would also allow for a critical look at the drill phenomenon and its potential relationship to the gang scene and real-life violence. Gang crime and violence have been around before the drill music scene (and before communicating such on YouTube). So a central question will have to be whether drill music and the spread of it through mainstream social media functions as a catalyst of the problem or merely brings it to the attention of the mainstream social media consumer.

## 4.1 Limitations and outlook

We acknowledge several limitations of this work. First, our work approached each lyric from a sentiment trajectory perspective. However, for applied purposes, it would also be useful to examine linguistic trajectories on the gang-level. For example, the two gangs Moscow17 and Zone2 have been in reported violent rivalry, including the killings of gang members and drill artists (Cobain, 2018). Future work could examine whether a trajectory analysis of both gangs' lyrics over time reflects shifts in violence that could be used as leading indicators for gang violence outbreaks.

Second, similar to studies of the language use of Chicago's gangs (Blevins et al., 2016; Chang et al., 2018), the drill music's language arguably differs from ordinary English, so that standard sentiment lexicons and even slang-specific ones fail to grasp the nuances of the content thoroughly. Whether or not a more nuanced approach affects the current findings is an empirical question for future research. A potentially promising approach might lie in the use of word embeddings to identify close semantic neighbours of out-of-vocabulary, highly context-specific idioms (e.g., Wijeratne et al., 2016).

Third, the current work looked at the linguistic modality of drill music and thereby ignored the visual modality (i.e. what can be seen in the video clips). Arguably, identical linguistic content will need to be interpreted differently depending on the contexts created by the visual content accompanying the lyrics. Similarly, our approach was agnostic to the auditory dimension of the lyrics and did not measure the intensity through which lyrics were performed. Future studies could look at multimodal approaches and seek to capture phenomena such as drill music in its full complexity.

## 4.2 Conclusion

Recent incidents have proposed the idea of a link between drill music and gang-related violence in London, UK. This work provided the first empirical insights into the language use of the London drill music scene. It can, therefore, lead to a better understanding of drill music language and help policymakers engage in evidence-based interventions regarding the alleged drill-violence nexus in future studies.

# 6 Appendix

**Appendix A:** Median shapes per sentiment trajectory cluster (standard sentiment lexicon).

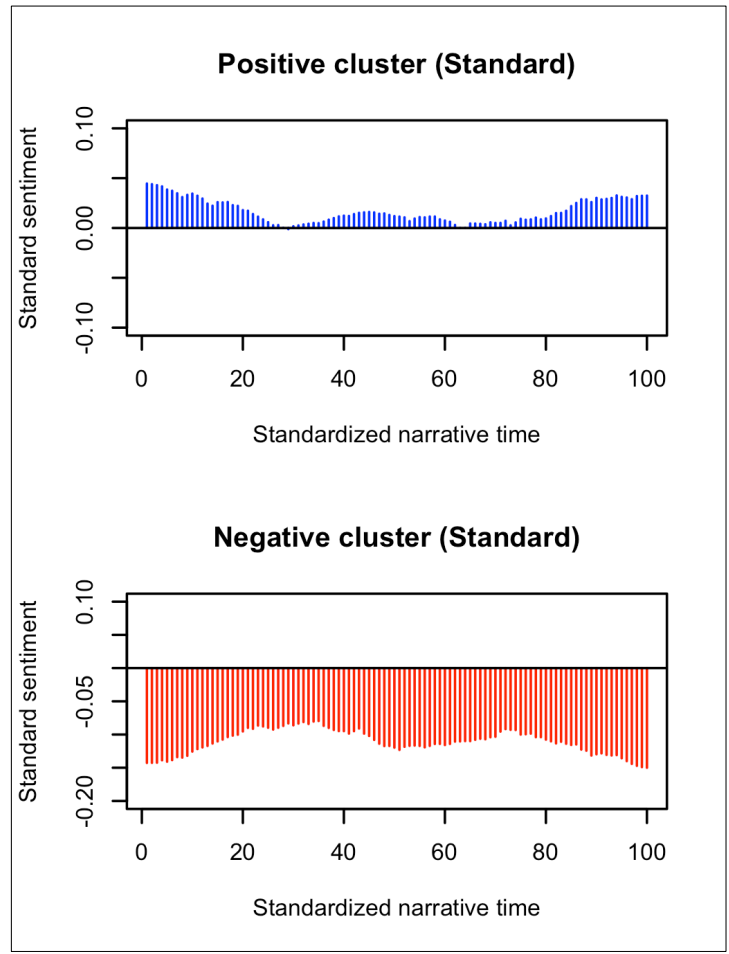

Figure A1: Median sentiment clusters using the standard sentiment lexicon.

**Appendix B:**
Median shapes per sentiment trajectory cluster (slang sentiment lexicon).

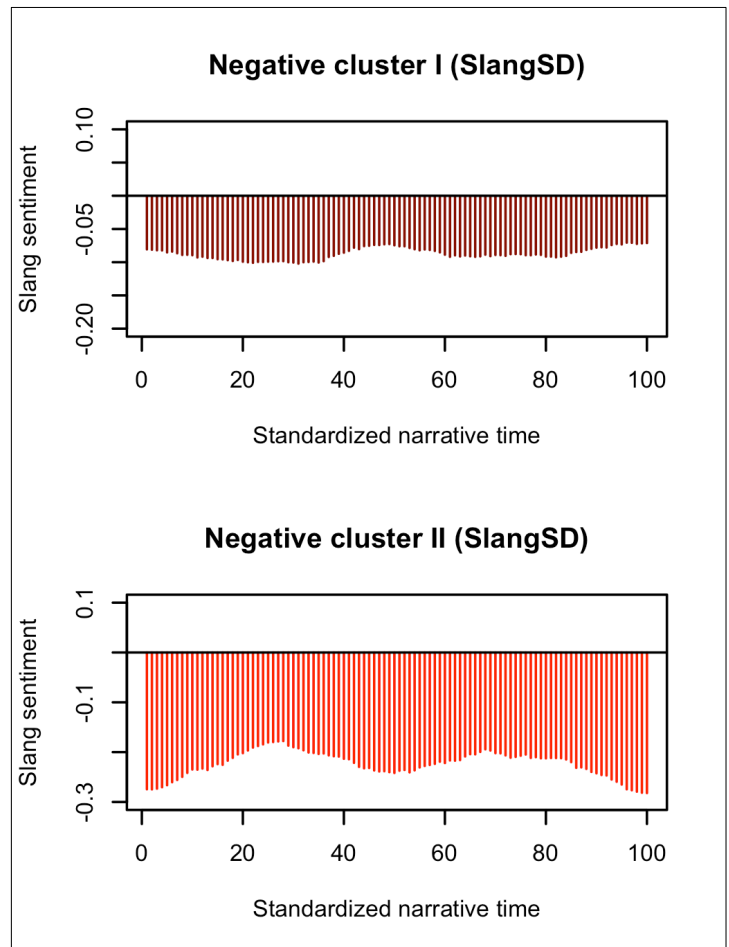

Figure B1: Median sentiment clusters using the slang sentiment lexicon.